\documentstyle[12pt]{article}
\begin{document}
\begin{titlepage}
\vspace{2.5cm}
\begin{centering}
{\LARGE{\bf Charge localization around disclinations in monolayer graphite}}\\
\bigskip\bigskip
S\'ergio Azevedo, Cl\'audio Furtado and Fernando Moraes\\
{\em Departamento de F\'{\i}sica\\
Universidade Federal de Pernambuco\\
50670-901 Recife, PE, Brazil}\\
\vspace{1cm}

\end{centering}
\vspace{1.5cm}

\begin{abstract}

Using a continuum model, we obtain qualitative results that imply charge 
localization around negative curvature disclinations (i.e. rings with more than 
6 Carbon atoms) in a graphite sheet. Conversely, it is found that positive 
curvature disclinations repel charge, independent of its sign.  
\end{abstract}

\hspace{0.8 cm}
\end{titlepage}
\def\carre{\vbox{\hrule\hbox{\vrule\kern 3pt
\vbox{\kern 3pt\kern 3pt}\kern 3pt\vrule}\hrule}}

\baselineskip = 18pt
\section{Introduction}
Although the electronic properties of elastic materials containing topological 
defects have been widely studied since the early 50's~\cite{Bar,Lan} in the 
context of deformation potential theory, there has been a recent revival of 
interest in such problems using both gauge field theory~\cite{Tei,Osi,Reb} and 
Riemann-Cartan geometry~\cite{Kat,Fur1,Fur2}. In this work we use the latter 
approach to study the effects of the curvature associated with disclinations in 
a monolayer of graphite.

Disclinations have a fundamental role in the formation of non-flat 
low-dimensional carbon structures such as fullerenes and graphene tubules among 
others. They are responsible for the local curvature that bends those structures
into their various shapes. This is easily seen by incorporating disclinations 
to  a graphite sheet. Graphite is the honeycomb-like, flat, two-dimensional
carbon network made solely of 6-membered rings. Rings made of a number of carbon
atoms other than 6 correspond to disclinations. They are formed, at least
conceptually, by a ``cut and glue" process characteristic of topological 
defects~\cite{Har}. For instance, to create a single 5-membered ring in an 
infinite graphite sheet cut out a wedge of angle $2\pi/6$ from the center of any
hexagon (so that one of the edges of the hexagon is removed) and identify the 
lose ends. The result is a conical graphitic structure with a  pentagonal ring 
at its (truncated) apex. Conversely, a 7-membered ring may be introduced by 
inserting a wedge of angle $2\pi/6$, adding an extra edge to one of the 
hexagons. Here, the result is a saddle-like structure with the heptagonal ring 
at the center. 

The  $n$-membered rings, with $n<6$, are an essencial part of the closed, 
positively curved,  fullerenes that are topologically equivalent to 
spheres~\cite{Uga} and of graphene tubules~\cite{Sai}. The rings with $n>6$ are 
indispensable to the open, negativelly curved, carbon networks~\cite{Fuj}. 
Toroidal~\cite{Joh} and helical~\cite{Dun} forms of carbon require both.

The electronic states of disclinated monolayer graphite were studied by Tamura 
and Tsukada~\cite{Tam} based on the connectivity of the atoms in the network, 
leaving out the effects of curvature. It is our aim here, to point out some of 
the curvature effects relevant for their discussion.

\section{The self-interaction}
In this section we obtain the self-energy and self-force of a point charge in a 
graphite sheet with a single disclination. We work in the continuum space limit,
where Riemannian geometry makes sense. Although a discrete geometrical approach
is also possible, along the lines of Regge calculus~\cite{Mis}, the properties 
we want to describe are easier to deduce this way. The ``cut and glue" process, 
in the two-dimensional continuum, produces  a disclination which is associated 
with a point with singular curvature. The interaction of a charged particle with
the defect, the deformation potential, results from the non-Euclidean metric of 
the defected medium and the modified topology which contribute with a 
self-energy to the charged particle~\cite{Fur1}.The corresponding 
two-dimensional space metric is given by~\cite{Kat}   
\begin{equation}
ds^2=dr^2+\frac{r^2}{p^2}d\theta^2,\,\,\,0\leq\theta\leq 2\pi,
\end{equation}
where $p$ measures the angular excess or deficit due to the disclination (since 
the total angle is $2\pi/p$ an angular excess or deficit of $\pm\lambda$ makes  
$2\pi/p=2\pi\pm\lambda$). The curvature tensor corresponding to this metric is 
zero everywhere except at the origin (where the defect is located) where it has 
a $\delta$-function singularity~\cite{Sok}. For $p>1$ the curvature is positive 
and for $0<p<1$ it is negative. $p=1$, zero curvature, corresponds to the 
absence of the defect.

In Reference~\cite{Fur1} we studied a charged particle in a disclinated 
three-dimensional medium. The self-energy  can be visualized as the energy 
associated with the distortion in the spacial distribution of the field lines of
a charge due to the defected background. This is similar to what happens to a 
charge near a conducting plane: the field lines are distorted due to the 
boundary condition on the plane and the charge is subjected to a force (which is
minus the gradient of its self-energy due to the distortion). We must be careful
here since, although the charge is confined to a two-dimensional surface, its 
field lines are distributed in three-dimensional space (three-dimensional 
electrodynamics: the Coulomb potential proportional to the inverse of the 
distance). So, only the field lines contained in the surface are affected by the
defect.  In what follows we take as the material medium (i.e. the graphite 
sheet) the surface $z=0$ and position the defect at the origin ($r=0$).

We basically follow the procedure used by Smith~\cite{Smi} in order to compute 
the self-force on a point charge in a background space of non-trivial topology. 
It goes as follows: we first solve Poisson's equation for a single charge in the
chosen background in order to find its Green function. The Green function is 
then renormalized by the extraction of its divergent part, namely the Green 
function in the absence of the defect. The self-energy is obtained from the 
coincidence limit of the renormalized Green function. Notice that, because we 
are dealing with three-dimensional electrodynamics (even though restricted to a 
two-dimensional surface) we need to work with the three-dimensional Green 
function.

We start then with Poison's equation
\begin{equation}
\nabla ^{2} G_{p}(\vec x, \vec x\,')=-4\pi\delta^{(3)}(\vec x - \vec x\,').
\end{equation}
In the defectless medium, $\nabla ^2$ is the usual three-dimensional Laplacian 
operator whereas in the disclinated medium it is given by~\cite{Fur1}
\begin{equation}
\nabla ^2 =\frac{\partial ^2}{\partial z^2}+\frac{1}{r}\frac{\partial}{\partial 
r}(r\frac{\partial}{\partial r})+\frac{p^2}{r^2}\frac{\partial
^2}{\partial\theta ^2},
\end{equation}
in cylindrical coordinates $z,r,\theta$, with $p\neq 1$ in the surface $z=0$ 
and 1 otherwise. That is, for $z\ne 0$ we still have the usual Laplacian but 
in the disclinated surface ($z=0$) it is modified by $p$. 

Defining
\begin{equation}
f(z)=1-[\Theta(z)+\Theta(-z)],
\end{equation}
where $\Theta(z)$ is the Heaviside step function such that $f(z)=1$ if $z=0$ 
and $f(z)=0$ if $z\ne 0$, we write the Green function as a sum of two parts
\begin{equation}
G_{p}(\vec x, \vec x\,')=f(z) G_{p}(\vec x, \vec x\,')+[1-f(z)]G_{p}(\vec x,
\vec x\,').
\end{equation}
The first term corresponds to the Green function in the $z=0$ surface while the 
second term corresponds to the Green function in the three-dimensional space 
minus this surface. Since the defect affects only the $z=0$ surface, $p=1$ in 
the second term. 

Now we need to regularize the Green function by subtraction of its defectless 
counterpart $G_{1}$, i.e.
\begin{equation}
G_{p}(\vec x, \vec x\,')_{ren}=G_{p}(\vec x, \vec x\,')-G_{1}(\vec x, \vec 
x\,')=f(z)[G_{p}(\vec x, \vec x\,')-G_{1}(\vec x, \vec x\,')].
\end{equation}
which is the three-dimensional renormalized Green function restricted to the 
$z=0$ surface. Since the next step is to take the coincidence limit, we can 
therefore use Smith's result for the three-dimensional renormalized Green 
function~\cite{Smi}:
\begin{equation}
G_{p}(\vec x, \vec x)_{ren}=\mathrel{\mathop{lim}\limits_{\vec x\,'\rightarrow 
\vec x}}[G_{p}(\vec x, \vec x\,')-G_{1}(\vec x, \vec x\,')]= \frac{1}{2\pi}\frac
{\kappa (p)}{r},
\end{equation}
where the numerical coefficient $\kappa (p)$ is given (here we use the notation 
of Reference \cite{Fur1} which differs from Reference \cite{Smi} by a factor of 
$2/\pi$)
\begin{equation}
\kappa (p)=2\int_{0}^{\infty}\frac{p\coth (px) - \coth (x)}{\sinh (x)}. 
\end{equation}
Since this result is independent of $z$ it already satisfies the restriction to 
the $z=0$ surface.

The self-energy of a point charge $q$ positioned at $\vec x$ is given 
by~\cite{Jac}
\begin{equation}
U(\vec x)=\frac{1}{2}\int\rho(\vec x-\vec x\,\,')\Phi(\vec x\,\,')d^{3}x'.
\end{equation} 
where
\begin{equation}
\rho(\vec x-\vec x\,')=q\delta ^{3}(\vec x- \vec 
x\,')=\frac{q}{r}\delta(r-r')\delta(\theta-\theta')\delta(z),
\end{equation}
and $\Phi(\vec x\,\,')=\frac{q}{\varepsilon}G(\vec x,\vec x\,')$ is the 
electrostatic potential at $\vec x\,'$ of the charge $q$ located at $\vec x$. 
Here $\varepsilon$ is the dielectric constant.

It follows that 
\begin{equation}
U(\vec x)=\frac{q^{2}}{2\varepsilon}G_{p}(\vec x,\vec x)_{ren}=\frac{q^{2}}{4\pi
\varepsilon}\frac{\kappa (p)}{r}
\end{equation}
and
\begin{equation}
\vec F(\vec x)=-\vec \nabla U(\vec x)=-\frac{\kappa (p)q^2}{4\pi \varepsilon
r^2}.
\end{equation}
It is interesting to notice here that this result is identical to the one of a 
point charge in the presence of a line disclination in three-dimensional medium.
In other words, the problem of  a point charge embedded in a two-dimensional 
surface with a point defect is the same as the charge in a three-dimensional 
medium with a line defect (except for the translational invariance along the $z$
direction). 

\begin{table}
\begin{center}
\begin{tabular}{|l|l|l|} \hline
$n$ & $p$ & $\kappa(p)$ \\ \hline
4 & 6/4 & 1.418 \\ \hline
5 & 6/5 & 0.5249 \\ \hline
6 & 1 & 0 \\ \hline
7 & 6/7 & -0.3351 \\ \hline
8 & 6/8 & -0.5622 \\ \hline
\end{tabular}
\end{center}
\caption{Numerical values for $\kappa(p)$ for selected $n$-membered rings.}
\end{table}

 Notice that $\kappa(1)=0$, meaning that the self-energy and self-force vanish 
in the absence of the defect. As mentioned in the Introduction, a five-membered 
Carbon ring corresponds to a disclination of deficit angle $2\pi /6$, whereas a 
7-membered ring corresponds to an excess angle of the same value, and so on. In 
Table 1 we list numerical values of $\kappa (p)$ for the different rings 
relevant to graphite. It is clear that for $p<1$ the coefficient $\kappa (p)<0$ 
implying an attractive self-force; i.e., charge of any sign will be attracted by
rings with less than 6 Carbon atoms. For $p>1$, we have that $\kappa (p)<0$ 
meaning that rings with more than 6 Carbon atoms repel charge. 

\section{Localization}
We study here the case $p<1$ which corresponds to a classical attractive 
self-force and which leads to quantum bound states. It was shown in the previous
section that we have the same situation of a charge in a three-dimensional 
medium with a line defect except that the translational invariance along the $z$
direction is broken. This enables us to use our previous results~\cite{Fur1}on 
the binding of charged particles to line disclinations. The two-dimensional 
Hamiltonian for a quasiparticle of charge $q=\pm e$ and effective mass $m^*$, 
moving in the strain field of a $p<1$ disclination is obtained from the 
Laplacian, Equation (3), and the self-energy, Equation (11):
\begin{equation}
-\frac{\hbar ^2}{2m^*}[\frac{1}{r}\frac{\partial}{\partial 
r}(r\frac{\partial}{\partial r})+\frac{p^2}{r^2}\frac{\partial ^2}{\partial 
\theta^2} ]-\frac{e^2}{4\pi \varepsilon}\frac{|\kappa (p)|}{r}
\end{equation}
with $0<p<1$. 

Schr\"{o}dinger's equation for this Hamiltonian is easily solvable. Writing the 
wavefunction as $R(r)e^{il\theta}$, where $l$ is an integer, the remaining 
radial equation,
\begin{equation}
r^{2}R''+rR'-(p^{2}l^{2}+\gamma r+\beta r^{2})R=0,
\end{equation}
where $\beta^{2}=-2m^{*}E/\hbar^{2}$, $\gamma =-(m^{*}e^{2}/2\pi \varepsilon 
\hbar^{2})|\kappa (p)|$ and $E$ is the energy, is a special case of the 
hypergeometric equation~\cite{Sea}. We find then the spacially localized 
eingenfunctions
\begin{equation}
R_{n,l}=C_{n,l}e^{-\beta r}(\beta r)^{|l|/\alpha}\,_{1}F_{1}(-n,2p|l|+1,\beta 
r),
\end{equation}
and the eigenenergies
\begin{equation}
E=-\frac{m^{*}e^{4}\kappa ^{2}(p)}{32\pi ^{2}\hbar ^{2} \varepsilon 
^{2}}\frac{1}{(n+\frac{1}{2}+p|l|)^2},
\end{equation}
where $C_{n,l}$ is a normalization constant, $_{1}F_{1}$ is the confluent 
hypergeometric function, $n=0,1,2, ...$ and $l=0,\pm 1, \pm 2, ...$.
 
\section{Concluding remarks}
In this work we obtained qualitative data on the dynamics of point charges near 
disclinations in a graphite sheet in a continuum model. We found that 
disclinations corresponding to rings with more than 6 Carbon atoms function as 
attractors to point charges. On the other hand, disclinations corresponding to 
rings with fewer than 6 Carbon atoms will repel the point charges. This is in 
sharp contrast with the topological analysis done by Tamura and 
Tsukada~\cite{Tam} who found that the five-membered ring is attractive and the 
seven-membered ring is repulsive to electrons. In their work they take into 
account only how atoms are connected to each other, and the effect of the 
curvature of the surface is not considered. In the approach of the present work 
we do the opposite, we are concerned exclusively with the curvature effect, not 
taking into account the connectivity. It is clear that some kind of 
interpolation between the two approaches is needed. Although our results are 
qualitative, and therefore not suitable for a numeric comparison, they strongly 
point out the importance of taking the curvature effects into account in 
calculations of the electronic structure of graphite sheets with disclinations.

\noindent
{\bf Acknowledgments}\\
\noindent
Most of this work was done while one of the authors (FM) was on a sabbatical 
leave at the Institute for Advanced Study, Princeton, NJ. We are grateful to 
CNPq for partial support of this work.


\begin{thebibliography}{99}
\bibitem{Bar}J. Bardeen and W. Shockley, Phys. Rev. {\bf 80}, 72 (1950).
\bibitem{Lan}R. Landauer, Phys. Rev. {\bf 82}, 520 (1951).
\bibitem{Tei}H. Teichler, Phys. Lett. A {\bf 87}, 113 (1981).
\bibitem{Osi}V. A. Osipov and S. E. Krasavin, J. Phys. C {\bf 7}, L95 (1995).
\bibitem{Reb}Y. T. Rebane, Phys. Rev. B {\bf 52}, 1590 (1995).
\bibitem{Kat}M. O. Katanaev and I. V. Volovich, Ann. Phys. (N.Y.) {\bf 216}, 1 
(1992).
\bibitem{Fur1}Claudio Furtado and Fernando Moraes, Phys. Lett. A {\bf 188}, 394 
(1994).
\bibitem{Fur2}C. Furtado, B. G. C. da Cunha, F. Moraes, E. R. Bezerra de Mello 
and V. B. Bezerra, Phys. Lett. A {\bf 195}, 90 (1994).
\bibitem{Har}W. F. Harris, Scient. Amer.  {\bf 237(6)}, 130 (1977).
\bibitem{Uga}D. Ugarte, Nature (London) {\bf 359}, 707 (1992).
\bibitem{Sai}M. Fujita, R. Saito, M. S. Dresselhaus, and G. Dresselhaus, Phys.
Rev. B {\bf45}, 13 834 (1992).
\bibitem{Fuj}Mitsutaka Fujita and Takahide Umeda, Phys. Rev. B {\bf 51}, 13778 
(1995).
\bibitem{Joh}J. K. Johnson {\em et al.}, Phys. Rev. B {\bf 50}, 17575 (1994).
\bibitem{Dun}B. I. Dunlap, Phys. Rev. B  {\bf 50}, 8134 (1994).
\bibitem{Mis}C. W. Misner, K. S. Thorne and J. A. Wheeler, Gravitation (Freeman,
San Francisco, 1973).
\bibitem{Sok}D.D. Sokolov and A.A. Starobinskii, Sov. Phys. Dokl. {\bf 22}, 312 
(1977).
\bibitem{Smi}A. G. Smith, in: The formation and evolution of cosmic strings, 
eds. G. W. Gibbons, S. W. Hawking and T. Vachaspati (Cambridge Univ. Press, 
Cambridge, 1990) pp. 263-292.
\bibitem{Jac}J. D. Jackson, Classical Electrodynamics (John Wiley, New York, 
1975).
\bibitem{Sea}James B. Seaborn, Hypergeometric Functions and their applications 
(Springer-Verlag, New York 1991).
\bibitem{Tam}Ryo Tamura and Masaru Tsukada, Phys. Rev. B {\bf 49}, 7697 (1994).
\end{thebibliography}
\end{document}